\documentclass[seceq]{jpsj2}
\title{Parity and Time-Reversal Breaking Effects on
Resonant X-Ray Scattering \\
at the Fe Pre-K-Edge in Magnetite}
\author{Jun-ichi \textsc{Igarashi}\thanks{E-mail:jigarash.mx.ibaraki.ac.jp} and Tatsuya \textsc{Nagao}$^{1}$}
\inst{Faculty of Science, Ibaraki University, Mito, Ibaraki 310-8512 \\
$^{1}$Faculty of Engineering, Gunma University, Kiryu, Gunma 376-8515}
\abst{We study noncentrosymmetric effects on resonant x-ray scattering 
(RXS) in magnetite.
The noncentrosymmetry at A sites in spinel structure makes
the $4p$ states strongly hybridize with the $3d$ states through neighboring 
oxygen $2p$ states, giving rise to 
the non-vanishing contribution of the dipole-quadrupole
($E$1-$E$2) process in the RXS spectra. 
We substantiate this observation by introducing a microscopic model
of a FeO$_4$ cluster with multiplets and the $4p$ band.
We show that the hybridization changes its sign between two kinds of A sites
and accordingly the local amplitude from the $E$1-$E$2 process changes 
its sign, resulting in non-vanishing RXS intensities at forbidden spots 
$(002)$ and $(006)$ in the pre-edge region in agreement with the experiment. 
A large dependence of the pre-edge intensity on the direction of the applied 
magnetic field is predicted as a consequence of breaking both centrosymmetry 
and time-reversal symmetry.
Furthermore we analyze the intensity difference between two opposite 
directions of the applied magnetic field at the allowed spot $(222)$ 
in connection with the experiment. We obtain the intensity difference
of a ``dispersion" form, which resembles the observed spectra at the Mn 
pre-$K$-edge in MnCr$_2$O$_4$ but is quite different from the observed one 
in magnetite.
Although the observed spectra are claimed to arise from ``magnetoelectric" 
amplitude, we argue that this claim has no ground.}
\kword{resonant x-ray scattering, magnetite,
local noncentrosymmetricity, E1-E2 
transition, magnetoelectric effect, XMCD, MnCr$_2$O$_4$}

\begin{document}
\maketitle

\section{\label{sect.1}Introduction}
Resonant x-ray scattering (RXS) has been widely used to investigate different
kinds of orders, such as charge, magnetic
and orbital orders in crystals,\cite{Hill1997,
Murakami1998-1,Murakami1998-2,Neubeck1999,Grenier2004}
since the strong resonance makes the signal sensitive to the ordered 
structure. The $K$-edge resonance is usually used in transition metals 
in order to observe signals at superlattice spots associated 
with the order parameter.
This is because the corresponding x-ray wavelength matches the period of 
long range orders, which is usually an order of atomic distance except 
for the case of long period.

The RXS amplitude is given by a sum of atomic amplitudes 
with appropriate phases. Each atomic amplitude is described by a second order 
process. One of the most dominant 
processes in transition-metal compounds is the dipole-dipole
($E$1-$E$1) process that 
the $1s$ electron is excited to the $4p$ states by absorbing 
x-ray and then 
the $4p$ electron is recombined with the $1s$-core hole by emitting x-ray. 
Since the $4p$ states are extended in space,
they are easily influenced by the electronic structure at neighbors
to the core-hole site; the $4p$ states are modulated by
the lattice distortion through the hybridization to 
neighboring oxygens,
giving rise to the signal at superlattice spots.
Therefore, the RXS signal at the superlattice spots arises from the variation 
of the $4p$ states in accordance with the long-range order, and provides
an indirect proof of the order which is usually constructed by $3d$ states.
Such a view has been confirmed by theoretical analyses
\cite{Elfimov1999, Benfatto1999,Taka1999}
in connection with the RXS experiment for LaMnO$_3$.\cite{Murakami1998-2}

There appear sometimes extra signals with energy below the 
$K$-edge, 
called the pre-edge signals, which will be mainly discussed in 
the present paper. Since the pre-edge energy is close to the energy exciting
an electron from the $1s$ state to the $3d$ states, the signal could be 
naturally interpreted as arising from the quadrupole-quadrupole
($E$2-$E$2) process in which the 
$1s$ electron is excited to the $3d$ states by absorbing x-ray and then one 
of $3d$ electrons is combined with the core hole by emitting 
x-ray. However,
the pre-edge signal could also be generated from the $E$1-$E$1
process, since the $p$-symmetric states with respect to the core-hole site
can be constructed from $3d$ states at neighboring transition-metal atoms.
\cite{Elfimov1999,Taka2000}
These two origins may be distinguished by different peak positions, that is,
the peak in the $E$2-$E$2 process is expected to be located at the region
around several eVs lower than that in the $E$1-$E$1 process, since
the relevant $3d$ states in the $E$2-$E$2 process is on the core-hole site,
and is strongly attracted by the core-hole potential.

The situation may become quite different when the centrosymmetry is locally 
broken. In such circumstances, the $4p$ states could hybridize with
the $3d$ states on the same site through the hybridization to neighboring 
oxygen $2p$ states, and thereby the dipole-quadrupole
($E$1-$E$2) process could contribute to
the pre-edge signals. Such a presence of the $E$1-$E$2 process has been
recognized by the experiment of K$_2$CrO$_4$\cite{Templeton1994} 
and by the numerical calculation for Ge.\cite{Elfimov2002}
It has been much debated for V$_2$O$_3$.\cite{Paolasini1999,Tanaka2002,
Joly2004,Lovesey2007} 
Furthermore, for magnetic materials,
the atomic amplitude of the $E$1-$E$2 process could depend on the direction 
of the local magnetic moment due to the spin-orbit interaction
(SOI), and thereby
the pre-edge signals could depend on the direction of magnetic moment.
Since the direction of magnetic moment could be controlled by applying the 
external magnetic field, we could observe such a dependence by changing 
the external magnetic field. Actually, such signals have been observed
\cite{Arima2005} and analyzed in a multiferroic 
system GaFeO$_3$,\cite{DiMatteo2006,Lovesey2007-2}
and also have been discussed in other 
situations.\cite{DiMatteo2005,Collins2007}

The pre-edge signals have also been observed in magnetite, Fe$_3$O$_4$,
at forbidden spots of scattering vectors $(002)$ and $(006)$.
\cite{Garcia2001,Kanazawa2002} 
In addition, the intensity difference with changing direction of the 
external magnetic field has been measured at spots $(222)$, 
$(333)$ and $(444)$.\cite{Kawata1995,Matsubara2005} 
The purpose of this paper is to analyze the pre-edge signals in magnetite
through a quantitative calculation of the spectra
and to elucidate the mechanism from a microscopic viewpoint.

\begin{figure}[h]
\begin{center}
\includegraphics[width=8.0cm]{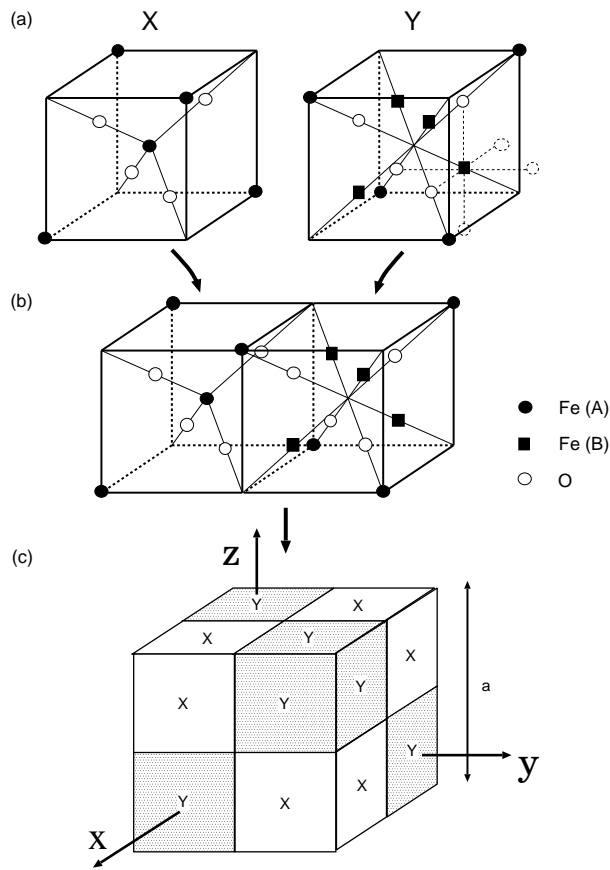}

\caption{Crystal structure of magnetite. The origin of coordinates
passes through the center of an Fe atom.
The unit cell contains 24 Fe atoms.
\label{structure}}

\end{center}
\end{figure}

The magnetite is the first magnetic material known to the mankind.
Its crystal structure is the inverse spinel, consisting of iron sites 
tetrahedrally surrounded by four oxygens (A sites) and those surrounded 
octahedrally by six oxygens (B sites), as shown in Fig.~\ref{structure}.
Since the centrosymmetry is locally broken at the
tetrahedral sites, 
those pre-edge signals are thought to be related to breaking
both centrosymmetry and time-reversal symmetry. 
Analyses based on the microscopic electronic structure, however,
have not been worked out yet.
We construct a definite model that the $4p$ states form an energy band 
with wide width and hybridize strongly with the $3d$ states through 
neighboring oxygen $2p$ states. The multiplet structures are taken into account
in the $3d^5$- and $3d^6$-configurations. Applying the resolvent formalism
\cite{Anderson1961} to this model,
we calculate the local electronic structure around the tetrahedron sites
and thereby the atomic amplitudes of RXS.

\begin{figure}[h]
\begin{center}
\includegraphics[width=8.0cm]{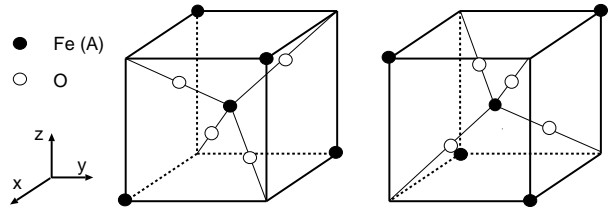}

\caption{Two types of tetrahedrons, $A_1$ (left) and $A_2$ (right).
\label{tetra} }

\end{center}
\end{figure}

It is important to recognize that there are two kinds of tetrahedron sites 
denoted as A$_1$ and A$_2$, that is, one is transformed into the other 
by space inversion with respect to the center of the tetrahedron 
(see Fig.~\ref{tetra}).
We find that the effective hybridization between the $4p$ and $3d$ states 
via oxygen $2p$ states changes its sign between the
A$_1$ and A$_2$ sites,
leading to a sign change in the atomic amplitude of the $E$1-$E$2 process.
This is a key point to explain how the pre-edge signals come out.
At spots $(002)$ and $(006)$, the contributions from the $E$1-$E$1 
and $E$2-$E$2 processes as well as Thomson scattering are canceled out
in the $\sigma-\sigma'$ channel, and that of the $E$1-$E$2 process 
at the A sites only survives in the total scattering 
amplitude.\cite{Com1} 
We obtain the pre-edge spectra as a single peak as a function 
of the photon energy,
in agreement with the experiment.\cite{Garcia2001,Kanazawa2002} 
Furthermore, we calculate explicitly the dependence on the direction of 
local magnetic moment in the atomic amplitude.
The depending parts are found about an order of magnitude smaller than 
non-dependent ones. From this calculation, we obtain the intensity differences
at spots $(002)$ and $(006)$ when the direction of the magnetization is 
changed from the $[1,-1,0]$ direction to the reverse, which shape looks 
like an "absorption" type as a function of photon energy. 
It may not be hard to detect these signals, 
since the magnitudes are about 1/5 to the corresponding pre-edge intensity 
peaks. 

We also analyze the dependence on the external magnetic 
field at spots $(222)$, $(333)$ and $(444)$ in connection 
with the recent experiment.\cite{Matsubara2005} 
These are allowed spots, where the Thomson scattering 
amplitude is dominant.
Focusing on the $E$1-$E$2 process at the A sites, 
we calculate the intensity
difference between two opposite directions of the applied magnetic field.
The difference arises from the interference between the
Thomson scattering 
amplitude and the $E$1-$E$2 amplitude. We show that the intensity differences
are nearly the same magnitude at both $(222)$ and $(333)$ spots 
but no difference at $(444)$, and that the shapes as a function of photon
energy look like a ``dispersion" form concentrated in the pre-edge region. 
In the experiment by 
Matsubara \textit{et al}.,\cite{Matsubara2005} however,
the intensity difference at the $(222)$ spot is distributed over the region
much wider than the pre-edge region and two orders of magnitude larger than 
that at $(333)$ and the calculated values. 
Also, the spectral shape is quite different from the ``dispersion" form. 
Matsubara \textit{et al}. claimed that 
the difference at $(222)$ 
arises from a ``magnetoelectric" amplitude, that is, a consequence of 
breaking both centrosymmetry and time-reversal symmetry. We argue that
this claim has no ground. Our finding of a ``dispersion" form for the intensity
difference has been observed in the experiment at the Mn pre-$K$-edge in 
MnCr$_2$O$_4$,\cite{Thesis} where Mn atoms are occupying at 
the A sites in spinel structure.
Since the pre-$K$-edge signal selects the contribution 
from the A sites, 
this experiment suggests that the calculated spectra correspond to 
the signal from the $E$1-$E$2 process.

This paper is organized as follows. In Sec. II, we briefly summarize 
fundamentals of magnetite. In Sec. III, we introduce 
the model Hamiltonian
and study the electronic structure around the 
A sites. In Sec. IV, 
we describe the excited states involving a $1s$-core hole by 
employing
the resolvent formalism. In Sec. V, we calculate the absorption spectra 
and discuss the x-ray magnetic circular dichroic (XMCD) spectra.
We calculate the RXS spectra in comparison with experiments.
The last section is devoted to concluding remarks.
The geometrical factors are summarized in Appendix.

\section{Fundamentals of magnetite}

The crystal structure of magnetite is the inverse spinel with the lattice 
constant $a_0=8.396 \ \textrm{\AA}$, 
as shown in Fig.~\ref{structure}.
The unit cell contains 24 iron atoms; 8 iron atoms are tetrahedrally coordinated
by 4 oxygens (A sites) and 16 iron atoms are octahedrally 
coordinated by 
6 oxygens (B sites). Note that two types of tetrahedrons exist within the A sites
(A$_1$ and A$_2$), as shown in Fig. \ref{tetra}.
It will be clarified in subsequent sections that the subtle difference in 
electronic structures between the A$_1$ and A$_2$ sites leads to important 
consequences on the RXS spectra. With disregarding small crystal distortion,
we have the position vectors of Fe atoms in the unit cell,
\begin{eqnarray}
{\bf r}_{{\rm A}_1} &:& \left(0,0,0\right), \left(0,\frac{1}{2},\frac{1}{2}
\right),
\left(\frac{1}{2},0,\frac{1}{2}\right), \left(\frac{1}{2},\frac{1}{2},0\right),
\nonumber\\
{\bf r}_{{\rm A}_2} &:& \left(\frac{1}{4},\frac{1}{4},\frac{1}{4}\right), 
\left(\frac{1}{4},\frac{3}{4},\frac{3}{4}\right), 
\left(\frac{3}{4},\frac{1}{4},\frac{3}{4}\right), 
\left(\frac{3}{4},\frac{3}{4},\frac{1}{4}\right),\nonumber\\
{\bf r}_{\rm B} &:& \left(\frac{5}{8},\frac{5}{8},\frac{5}{8}\right), 
                \left(\frac{5}{8},\frac{1}{8},\frac{1}{8}\right), 
                \left(\frac{1}{8},\frac{5}{8},\frac{1}{8}\right), 
                \left(\frac{1}{8},\frac{1}{8},\frac{5}{8}\right),\nonumber\\
             && \left(\frac{5}{8},\frac{7}{8},\frac{7}{8}\right), 
                \left(\frac{5}{8},\frac{3}{8},\frac{3}{8}\right), 
                \left(\frac{1}{8},\frac{7}{8},\frac{3}{8}\right), 
                \left(\frac{1}{8},\frac{3}{8},\frac{7}{8}\right),\nonumber\\
             && \left(\frac{7}{8},\frac{5}{8},\frac{7}{8}\right), 
                \left(\frac{7}{8},\frac{1}{8},\frac{3}{8}\right), 
                \left(\frac{3}{8},\frac{5}{8},\frac{3}{8}\right), 
                \left(\frac{3}{8},\frac{1}{8},\frac{7}{8}\right),\nonumber\\
             && \left(\frac{7}{8},\frac{7}{8},\frac{5}{8}\right), 
                \left(\frac{7}{8},\frac{3}{8},\frac{1}{8}\right), 
                \left(\frac{3}{8},\frac{7}{8},\frac{1}{8}\right), 
                \left(\frac{3}{8},\frac{3}{8},\frac{5}{8}\right).
\nonumber \\
\label{eq.position}
\end{eqnarray}

Iron atoms at the A sites are nominally Fe$^{3+}$, 
while those at the B sites
are a $1:1$ mixture of Fe$^{3+}$ and Fe$^{2+}$. Note that Fe$^{3+}$ atoms are
in the $3d^5$-configuration with the spin angular momentum $S=5/2$,
and that Fe$^{2+}$ atoms are in the $3d^6$-configuration with $S=2$,
according to the Hund rule. The hybridization between the $3d$ states and
oxygen $2p$ states may change but slightly the situation.
The local magnetic moments are ferromagnetically aligned 
within the individual A and B sites, 
while those at the A and B sites are
antiferromagnetically aligned. As a result, a net magnetization remains finite,
that is, the magnetite is a ferrimagnet at low temperatures. 
The Curie temperature is as high as around 850 K.
In addition, a metal-insulator transition so called Verway\cite{Verway1939}
transition takes place around $T=120$ K. 
This may be related to charge and orbital orders, which study is
outside our scope.\cite{Leonov2006,Nazarenko2006,Uzu2006}

\section{Electronic Structures around the A sites}

In this section we focus on the electronic structure 
around the A sites
which have no centrosymmetry. In particular, we are interested in
the excited states having one $1s$ core hole and one $4p$ electron
in accordance with the $E$1 process and those having one $1s$ core hole
and one more electron in the $3d$ states in accordance with the $E$2 process.

\subsection{Crystal electric field}

We start by examining the crystal electric field (CEF)
to look for 
noncentrosymmetric effects.
Let charge $q$ be placed at the apexes of a tetrahedron. 
The electrostatic potential $\phi(x,y,z)$ is expanded around the center as
\begin{equation}
 \phi(x,y,z)=\frac{4q}{r_0}
            \mp \frac{20}{\sqrt{3}}\frac{q}{r_0^4}xyz \nonumber
    -\frac{35}{9}\frac{q}{r_0^5}\left(x^4+y^4+z^4-\frac{3}{5}r^4\right),
\label{eq.cef}
\end{equation}
where $r=\sqrt{x^2+y^2+z^2}$ with $r_0$ being the distance between the origin 
and the apexes. 
The last term is well known to represent a split of the energy level of 
$3d$ states.
The second term, which is usually neglected, gives rise to a hybridization
between $3d$ and $4p$ states. This coupling comes out because of 
noncentrosymmetry, but it is much smaller than the same type of coupling 
arising from the hybridization between the $3d$ and oxygen 
$2p$ states and between the $4p$ and $2p$ states.
The sign $-(+)$ of the coupling is taken for the
$A_1(A_2)$ sites.

\subsection{Effective hybridization between the $4p$ and $3d$ states}

Now we discuss how the $4p$ states could hybridize with the $3d$ states
in the absence of centrosymmetry.
Let $3d$ wavefunctions be 
$\psi^{3d}_{x^2-y^2}$, $\psi^{3d}_{3z^2-r^2}$,
$\psi^{3d}_{yz}$, $\psi^{3d}_{zx}$, and $\psi^{3d}_{xy}$,
and $4p$ wavefunctions be $\psi^{4p}_{x}$, $\psi^{4p}_{y}$, 
and $\psi^{4p}_{z}$.
They are all real and normalized, and have symmetries described in 
the subscript.
Each state could hybridize with a state constructed from a linear combination 
of oxygen $2p$ wavefunctions at apexes.
These oxygen wavefunctions have the same symmetry as their partner of
hybridization, which are denoted as 
$\psi^{2p}_{x^2-y^2}$, $\psi^{2p}_{3z^2-r^2}$,
$\psi^{2p}_{yz}$, $\psi^{2p}_{zx}$, $\psi^{2p}_{xy}$,
$\psi^{2p}_{x}$, $\psi^{2p}_{y}$, and $\psi^{2p}_{z}$.
Using the Slater-Koster two-center integrals given in Table \ref{table.1},
we evaluate the strength of hybridization between the $3d$ and $2p$ states
and between the $4p$ and $2p$ states, 
\begin{eqnarray}
t_{E}^{3d-2p}&=&
\langle \psi^{3d}_{x^2-y^2}|H_{\rm hyb}^{3d-2p}|\psi^{2p}_{x^2-y^2}\rangle \nonumber \\
&=&\langle \psi^{3d}_{3z^2-r^2}|H_{\rm hyb}^{3d-2p}|\psi^{2p}_{3z^2-r^2}\rangle
= 1.34\,\, {\rm eV} \label{eq.3d2p1}\\ 
t_{T_2}^{3d-2p}&=&\langle \psi^{3d}_{yz}|H_{\rm hyb}^{3d-2p}|\psi^{2p}_{yz}
\rangle 
=\langle \psi^{3d}_{zx}|H_{\rm hyb}^{3d-2p}|\psi^{2p}_{zx}\rangle
\nonumber \\ 
&=&\langle \psi^{3d}_{xy}|H_{\rm hyb}^{3d-2p}|\psi^{2p}_{xy}\rangle
=\mp 2.33\,\, {\rm eV}, \quad {\rm for\,\, A_1(A_2)} ,\label{eq.3d2p2}\\
t^{4p-2p}&=&\langle \psi^{4p}_{x}|H_{\rm hyb}^{4p-2p}|\psi^{2p}_{x}\rangle
=\langle \psi^{4p}_{y}|H_{\rm hyb}^{4p-2p}|\psi^{2p}_{y}\rangle
\nonumber \\
&=&\langle \psi^{4p}_{z}|H_{\rm hyb}^{4p-2p}|\psi^{2p}_{z}\rangle
= -4.36 \,\, {\rm eV},
\label{eq.4p2p}
\end{eqnarray}
where $H_{\rm hyb}^{3d-2p}$ and $H_{\rm hyb}^{4p-2p}$ are the hybridization
energies between the $3d$ and $2p$ states and between $4p$ and $2p$ states,
respectively. The sign $-(+)$ in eq.~(\ref{eq.3d2p2}) corresponds to the A$_1$ 
(A$_2$) sites. Note that $\psi^{2p}_{yz}$ and $\psi^{2p}_{x}$
are not identical but have a finite overlap.
The same is true for 
$\psi^{2p}_{zx}$ and $\psi^{2p}_{y}$ and for
$\psi^{2p}_{xy}$ and $\psi^{2p}_{z}$, respectively.
The overlap is evaluated as
\begin{equation}
S=\langle \psi^{2p}_x|\psi^{2p}_{yz}\rangle =
\langle \psi^{2p}_y|\psi^{2p}_{zx}\rangle =
\langle \psi^{2p}_z|\psi^{2p}_{xy}\rangle = -0.748.
\end{equation}
Needless to say, these values depend on the phase of wavefunctions
constructed from oxygen 2p orbitals, but the effective hybridization
between the $4p$ and $3d$ states are independent of the phase,
since it is proportional to 
$\langle \psi^{4p}_{x}|H_{\rm hyb}^{4p-2p}|\psi^{2p}_{x}\rangle 
\langle \psi^{2p}_x|\psi^{2p}_{yz}\rangle 
\langle \psi^{2p}_{yz}|H_{\rm hyb}^{3d-2p}|\psi^{3d}_{yz}\rangle$.
Its sign is opposite between the $A_1$ and $A_2$ sites.
This corresponds to the sign change of the second term of 
eq.~(\ref{eq.cef}) in the CEF.

\subsection{Hamiltonian for a FeO$_4$ cluster}

Now that the $4p$ states could hybridize with the $3d$ states
through oxygen $2p$ states, we include oxygen states into
our model, in addition to the $1s$, $3d$, and $4p$ states, 
in order to describe
the electronic structure around the A sites. For this reason, 
we consider the Hamiltonian of a FeO$_4$ cluster at the A sites,
\begin{equation}
 H = H^{3d} + H^{2p} + H_{\textrm{hyb}}^{3d-2p} + H^{1s} + H^{4p} 
 + H_{\textrm{hyb}}^{4p-2p}, \label{eq.Ham}
\end{equation}
where 
\begin{eqnarray}
 H^{3d} & = & \sum_{m\sigma}E_m^{d}d^{\dagger}_{m\sigma}d_{m\sigma}
  + \frac{1}{2}\sum_{\nu_{1}\nu_{2}\nu_{3}\nu_{4}}
  g\left(\nu_{1}\nu_{2};\nu_{3}\nu_{4}\right)d_{\nu_{1}}^{\dagger}
  d_{\nu_{2}}^{\dagger}d_{\nu_{4}}d_{\nu_{3}} \nonumber\\
& + & \zeta_{3d}\sum_{mm'\sigma\sigma'}
 \langle m\sigma|{\bf l}\cdot{\bf s}|m'\sigma'\rangle
  d^{\dagger}_{m\sigma}d_{m'\sigma'}
 +({\bf H}_{\rm xe}+{\bf H}_{\rm ext})\cdot \sum_{m\sigma\sigma'}
   ({\bf s})_{\sigma\sigma'}d^{\dagger}_{m\sigma}d_{m\sigma'},
\label{eq.H3d}\\
 H^{2p} &=& \sum_{m\sigma}E^{p}p^{\dagger}_{m\sigma}p_{m\sigma},\\
 H_{\rm hyb}^{3d-2p} &=& \sum_{m\sigma}t_{m}^{3d-2p}
   d_{m\sigma}^{\dagger}p_{m\sigma}+{\rm H.c.}, \\
 H^{1s} &=& \epsilon_{1s}\sum_{\sigma}s^{\dagger}_{\sigma}s_{\sigma},\\
 H^{4p} &=& \sum_{{\bf k}\eta\sigma}\epsilon_{4p}({\bf k})
   p'^{\dagger}_{{\bf k}\eta\sigma} p'_{{\bf k}\eta\sigma},\\
 H_{\rm hyb}^{4p-2p} &=& t^{4p-2p}
 \sum_{\eta\sigma} p'^{\dagger}_{\eta\sigma} p_{\eta\sigma}+{\rm H.c.}, 
\end{eqnarray}

The $H^{3d}$ describes the energy of $3d$ electrons, where
$d_{m\sigma}$ represents an annihilation operator of a $3d$ electron 
with spin $\sigma$ and symmetry $m$ ($=x^2-y^2,3z^2-r^2,yz,zx,xy$)
at the center.
The $3d$ energy level $E^{d}_{m}$ is split by the CEF energy
$10Dq$. The second term in eq.~(\ref{eq.H3d}) represents the intra-atomic 
Coulomb interaction with the interaction matrix element 
$g\left(\nu_{1}\nu_{2};\nu_{3}\nu_{4}\right)$ in terms of $F^{0}$, $F^{2}$, 
and $F^{4}$ ($\nu$ stands for $\left(m,\sigma\right)$).
The third term in eq.~(\ref{eq.H3d}) represents the SOI 
for $3d$ electrons with the SOI coupling $\zeta_{3d}$.
We evaluate atomic values of $F^2$, $F^4$, and 
$\zeta_{3d}$ using the 
wavefunctions within the Hartree-Fock (HF) 
approximation,\cite{Cowan1981}
and multiply $0.8$ to these atomic values with taking account 
of the slight screening effect.\cite{deGroot1990} 
On the other hand, we  multiply 0.25 to the atomic value 
for $F^0$, since it is known that $F^{0}$ is considerably screened 
by solid-state effects.
The last term in eq.~(\ref{eq.H3d}) describes the energy due to
the exchange interaction from neighboring Fe atoms and the Zeeman energy
with the external field, where
$({\bf s})_{\sigma\sigma'}$ represents the matrix element of
the spin operator of the $3d$ electrons. 
The exchange field ${\bf H}_{\rm xc}$ here has a dimension of energy,
and is an order $k_{\rm B}T_{c}$ with $T_{c}=850$ K.
The external field ${\bf H}_{\rm ext}$ is assumed to be much smaller than 
${\bf H}_{\rm xc}$ but to be larger than the magnetic anisotropy energy.
Therefore it has a role to align the magnetization to the field.

The $H^{2p}$ represents the energy of oxygen $2p$ electrons, where 
$p_{m\sigma}$ is the annihilation operator of the state 
$|\psi^{2p}_{m}\rangle$ with spin $\sigma$.
The Coulomb interaction is neglected in oxygen 2p states. 
The $H^{3d-2p}_{\rm hyb}$ represent the mixing energy between the $3d$ and
$2p$ states, where $t^{3d-2p}_m$ is the matrix element given by 
eqs.~(\ref{eq.3d2p1}) and (\ref{eq.3d2p2}).
The energy of the $2p$ level relative to the $3d$ levels is determined from 
the charge-transfer energy $\Delta$ defined by 
$\Delta=E^{d}-E^{p}+15U(3d^6)-10U(3d^5)$ with $E^d$ being an average of 
$E^d_m$. Here $U(3d^6)$ and $U(3d^5)$ are the multiplet-averaged $d$-$d$ 
Coulomb interaction in the $3d^6$ and $3d^5$ configurations, 
which are defined by $U=F^{0}-\left(2/63\right)F^{2}-\left(2/63\right)F^{4}$
with $F^{0}$, $F^{2}$, and $F^{4}$.

The last three terms are added to the Hamiltonian in accordance with the
excitation of the core electron.
The $H^{1s}$ represents the energy of the $1s$ electrons, where
$s_{\sigma}$ is an annihilation operator of the $1s$ state.
The $H^{4p}$ represents the energy of the $4p$ states, which form a band 
with energy $\epsilon_{4p}({\bf k})$. 
The $H^{4p-2p}_{\rm hyb}$ represents the hybridization between the $4p$
and oxygen $2p$ states, where $p'_{\eta\sigma}$ is the annihilation 
operator of $4p$ electron with symmetry $\eta=x,y$, and $z$, and
$p'_{\eta\sigma}=(1/\sqrt{N_0})\sum_{\bf k} p'_{{\bf k}\eta\sigma}$
($N_0$ is the number of ${\bf k}$).
This expression could be changed into a form that $4p$ states hybridize with
oxygen states symmetrized as $yz$, $zx$ and $xy$:
\begin{equation}
 H_{\rm hyb}^{4p-2p} = \tilde{t}^{4p-2p}
 \sum_{\eta\sigma} p'^{\dagger}_{\eta\sigma} p_{m\sigma}+{\rm H.c.}, 
\end{equation}
with $m=yz$ corresponding to $\eta=x$ and so on. Here the matrix element 
$t^{4p-2p}$ is renormalized as $\tilde{t}^{4p-2p}\equiv t^{4p-2p}S$.
We do not explicitly consider the Coulomb interactions 
between the core hole and the $4p$ and $3d$ electrons,
but we could take account of the main effects by adjusting the energy 
separation between $3d$ level and $\epsilon_{1s}$,
since the Slater integrals responsible to the exchange interaction is
rather small, $G^2(1s,3d)=0.058$ eV.

Table \ref{table.1} lists the parameter values used in this paper,
which are consistent with the values in the 
previous calculations.\cite{Chen2004}

\begin{table}[tb]
\caption{Parameter values for a FeO$_{4}$ cluster in 
the $3d^5$ configuration, measured in units of eV.}
\label{table.1}
\begin{tabular}{lrlr}
\hline
$F^0(3d,3d)$ & 6.39 & $(pd\sigma)_{2p,3d}$ & -1.9 \\
$F^2(3d,3d)$ & 9.64 & $(pd\pi)_{2p,3d}$ & 0.82 \\
$F^4(3d,3d)$ & 6.03 & $(pp\sigma)_{2p,4p}$ & 3.5 \\
$\zeta_{3d}$ & 0.059 & $(pp\pi)_{2p,4p}$ & -1.0 \\
$\Delta$ & 3.5 & $10 Dq$ & -0.7 \\
\hline
\end{tabular}
\end{table}

\subsection{Lowest energy state at the A sites}

Iron atoms at the A sites are nominally the Fe$^{3+}$ ($3d^5$) configuration.
The hybridization between the $3d$ states and oxygen $2p$ states makes it
mix with the $3d^{6}\underline{L}$ configuration, where $\underline{L}$ 
indicates the presence of a hole in the ligand oxygen orbitals.
Preparing $2352$ bases in the $3d^5+3d^6\underline{L}$ configuration, 
we represent the Hamiltonian $H_{3d}+H_{2p}+H_{\rm hyb}^{3d-2p}$
for ${\bf H}_{\rm xc}$ along the $z$ axis. Diagonalizing numerically 
the Hamiltonian matrix, we obtain the spin moment $S=2.40$ and the orbital
moment $L=0.0036$ in the lowest energy state. These values deviate slightly
from $S=5/2$ and $L=0$ in the lowest state of the $3d^5$ configuration.
Note that these values are insensitive to the magnitude and direction of 
${\bf H}_{\rm xc}$.
The weight of the $3d^5$ and $3d^6\underline{L}$ configurations 
are obtained as $0.795$ and $0.205$.

\section{Excited states relevant to the $K$ edge RXS}

\subsection{Resolvent formalism}

We use the resolvent formalism in order to describe the excited states 
containing a $1s$ core hole and a $4p$ electron. It is defined by
\begin{equation}
 G(z) = \left[z-H_0-V\right]^{-1},
\end{equation}
where
\begin{eqnarray}
 H_0 &=& H^{3d} + H^{2p} + H^{3d-2p} + H^{1s} + H^{4p}, \\
 V &=& H_{\rm hyb}^{4p-2p} .
\end{eqnarray}

Now let $|\beta\rangle$ and $|\gamma\rangle$ be eigenstates of $H_0$ 
with energies $E_{\beta}$ and $E_{\gamma}$ in the configuration of 
$3d^5 + 3d^6\underline{L}$ and in the $3d^6$ configuration, respectively.
These energies are defined from the ground state energy.
The excited states containing a pair of a $4p$ electron and a $1s$-core hole,
which is created by the $E$1 transition, may be given by
$p'^{\dagger}_{\eta\sigma}s_{\sigma}|\beta\rangle$.
Also the excited states caused by the $E$2 transition may be given by
$|c\rangle = s_{\sigma}|\gamma\rangle$.
States $|\beta\rangle$'s span the space of 2352 dimensions,
and $|\gamma\rangle$'s span the space of 210 dimensions.
Noting that the $1s$ hole and the $4p$ electron are coupled to 
$3d-2p$ electrons only through $V$, we have
\begin{eqnarray}
 \left[G(z)\right]_{\sigma\gamma,\sigma\gamma'} &\equiv&
 \langle\gamma|s^{\dagger}_{\sigma}(z-H)^{-1}s_{\sigma}|\gamma'\rangle 
 \nonumber \\
 &=& \left[(z-E_{\gamma}+\epsilon_{1s})\delta_{\gamma,\gamma'}
  - \sum_{\eta\beta}V_{\gamma,\eta\beta}G_{0}(z-E_{\beta})V_{\eta\beta,\gamma'}
  \right]^{-1},
\label{eq.ggg}
\end{eqnarray}
where
\begin{eqnarray}
 V_{\eta\beta,\gamma}
 &=& \langle\beta| p'_{\eta\sigma} H^{4p-2p}_{\rm hyb}|\gamma\rangle,\\
 G_{0}(z) &=& \frac{1}{N_0}\sum_{\bf k}\frac{1}
 {z-\epsilon_{4p}({\bf k})+\epsilon_{1s}+i\Gamma}, 
\label{eq.g0}
\end{eqnarray}
with $\Gamma$ being the core-hole life-time broadening width.
The right hand side of eq.~(\ref{eq.ggg}) means the inverse of
the matrix whose components are written inside the brace.
The inversion of the matrix is numerically carried out.

Once we know $[G(z)]_{\sigma\gamma,\sigma\gamma'}$, we immediately obtain
other components of the Green function,
\begin{eqnarray}
 [G(z)]_{\sigma\eta\beta,\sigma\gamma} &\equiv&
  \langle\beta|s^{\dagger}_{\sigma} p'_{\eta\sigma} 
  G(z)s_{\sigma}|\gamma\rangle\nonumber \\
 &=& G_0(z-E_{\beta})\sum_{\gamma'}V_{\eta\beta,\gamma'}
 [G(z)]_{\sigma\gamma',\sigma\gamma}, \label{eq.g4p-3d}\\
 G(z)_{\sigma\gamma,\sigma\eta\beta} &\equiv&
  \langle\gamma|s^{\dagger}_{\sigma} G(z) 
  p'^{\dagger}_{\eta\sigma} s_{\sigma} |\beta\rangle \nonumber\\
 &=&\sum_{\gamma'}[G(z)]_{\sigma\gamma,\sigma\gamma'}
 V_{\gamma',\eta\beta}G_0(z-E_{\beta}), \label{eq.g3d-4p}\\
 G(z)_{\sigma\eta\beta,\sigma\eta'\beta'} &\equiv&
 \langle\beta| s^{\dagger}_{\sigma} p'_{\eta\sigma} G(z) 
 p'^{\dagger}_{\eta'\sigma} s_{\sigma}|\beta\rangle , \nonumber \\
 &=& G_0(z)\delta_{\eta,\eta'}\delta_{\beta,\beta'}\nonumber\\
 &+& G_0(z-E_{\beta})\sum_{\gamma\gamma'} V_{\eta\beta,\gamma}
 [G(z)]_{\sigma\gamma,\sigma\gamma'}V_{\gamma',\eta'\beta'}G_0(z-E_{\beta'}).
\label{eq.g4p-4p}
\end{eqnarray}
The Green function is diagonal with the $\sigma$ variable.
It should be noted here that the off-diagonal components, 
$[G(z)]_{\sigma\eta\beta,\sigma\gamma}$ and 
$[G(z)]_{\sigma\gamma,\sigma\eta\beta}$ given by eqs.~(\ref{eq.g4p-3d})
and (\ref{eq.g3d-4p}), change their signs between the $A_1$ and $A_2$ sites,
in accordance with the change of the effective coupling between $4p$ 
and $3d$ states. In eq.~(\ref{eq.g4p-4p}), the last term  could not appear
at the centrosymmetric sites, since it arises from the effective coupling
between the $4p$ and $3d$ states which are not allowed within the same site. 
Note that, if a larger size of cluster is considered, the $p$-symmetric states 
could couple to such "3d" states through neighboring iron sites.

Among many $|\beta\rangle$'s in the $3d^5+3d^6\underline{L}$ configuration,
the lowest energy state $|\beta_0\rangle$ is taken into account 
in the following calculation. This may be justified when the presence
of the pair of $4p$ electron and $1s$-core hole could not modify the
$3d$ states through the Coulomb interaction.
This observation simplifies greatly the analysis of the K-edge RXS
in the next section.

\section{X-ray absorption and scattering near the K-edge of Iron}

\subsection{Transition matrix elements}

We need to consider two processes around the $K$-edge; one is the $E$1 process
that the $1s$ core electron is excited to the $4p$ states,
and the other is the $E$2 process that the $1s$-core electron is excited to
the $3d$ states. These processes may be represented by transition operators,
\begin{equation}
 T_{\eta\sigma}^{\rm E1}(j) = M_1 p'^{\dagger}_{\eta\sigma}s_{\sigma}, \quad 
 T_{m\sigma}^{\rm E2}(j) = M_2 d^{\dagger}_{m\sigma}s_{\sigma},
\end{equation}
where $\eta$ ($=x,y,z$) and $m$ ($=x^2-y^2,3z^2-r^2,yz,zx,xy$) are connected to 
the polarization of the incident photon.
The annihilation and creation operators are defined with respect to
the $j$th iron site. Since the $1s$ state is well localized around the iron site,
$M_1$ and $M_2$ are evaluated by using atomic wavefunctions. 
We have 
\begin{eqnarray}
 M_1 &=& iq\int\langle\psi^{4p}_{x}|x|\psi^{1s}\rangle {\rm d}^{3}r 
 \nonumber\\
     &=& iq \frac{1}{\sqrt{3}}
 \int_0^{\infty}r^3R_{4p}(r)R_{1s}(r){\rm d}r
      =i\,4.46 \times 10^{-3} ,\\
 M_2 &=& -q^2\int\langle\psi^{3d}_{zx}|(zx/2)|\psi^{1s}\rangle{\rm d}^{3}r
 \nonumber\\ 
     &=& -q^2 \frac{1}{2\sqrt{15}}
 \int_0^{\infty}r^4R_{3d}(r)R_{1s}(r){\rm d}r
 = -4.07\times 10^{-4}, 
\end{eqnarray}
where $R_{1s}(r)$, $R_{3d}(r)$, and $R_{4p}(r)$ are radial wavefunctions 
of the $1s$, $3d$, and $4p$ states, respectively, 
which are calculated within 
the HF approximation.\cite{Cowan1981}
We have inserted $q \sim 3.6\times 10^{8}$ ${\rm cm}^{-1}$ for the x-ray 
wavenumber, which corresponds to the K-edge energy $7.12$ keV.

\subsection{Absorption and XMCD spectra}

Although our main concern in this paper is the RXS spectra,
we briefly discuss the absorption spectra for looking over the whole $K$-edge 
region.

The absorption coefficient may be given by a sum of 
contributions from each site, since the $1s$ state is well localized 
at one atomic site. In general, it is decomposed into the contributions 
of the $E$1-$E$1, $E$1-$E$2, $E$2-$E$1, and $E$2-$E$2 processes:
\begin{eqnarray}
 A^{11}_{\eta\eta}(\omega) &=& \sum_{j,n,\sigma}\langle g|
 T^{E1\dagger}_{\eta\sigma}(j)|n\rangle\langle n|T^{E1}_{\eta\sigma}(j)
 |g\rangle \delta(\omega-E_n+E_g), \label{eq.a11}\\
 A^{12}_{\eta m}(\omega) &=& \sum_{jn,\sigma}\langle g|
 T^{E1\dagger}_{\eta\sigma}(j)|n\rangle\langle n|T^{E2}_{m\sigma}(j)
 |g\rangle \delta(\omega-E_n+E_g), \\
 A^{21}_{m \eta}(\omega) &=& \sum_{j,n,\sigma}\langle g|
 T^{E2\dagger}_{m\sigma}(j)|n\rangle\langle n|T^{E1}_{\eta\sigma}(j)
 |g\rangle \delta(\omega-E_n+E_g), \\
 A^{22}_{m m}(\omega) &=& \sum_{j,n,\sigma}\langle g|
 T^{E2\dagger}_{m\sigma}(j)|n\rangle\langle n|T^{E2}_{m\sigma}(j)
 |g\rangle \delta(\omega-E_n+E_g), \label{eq.a22}
\end{eqnarray}
where $|g\rangle$ and $|n\rangle$ represent the ground and
excited states
of the system with energy $E_g$ and $E_n$.
For example, when the x-ray is traveling along the z-direction with the 
polarization vector along the x-direction, we need to set $\eta=x$ and $m=zx$. 
To include the life-time broadening of the core level, we replace the
$\delta$-function in eqs.~(\ref{eq.a11})-(\ref{eq.a22}) by the Lorentzian
function with the full width of half maximum $2\Gamma=1.6$ eV.

\begin{figure}[h]
\begin{center}
\includegraphics[width=8.0cm]{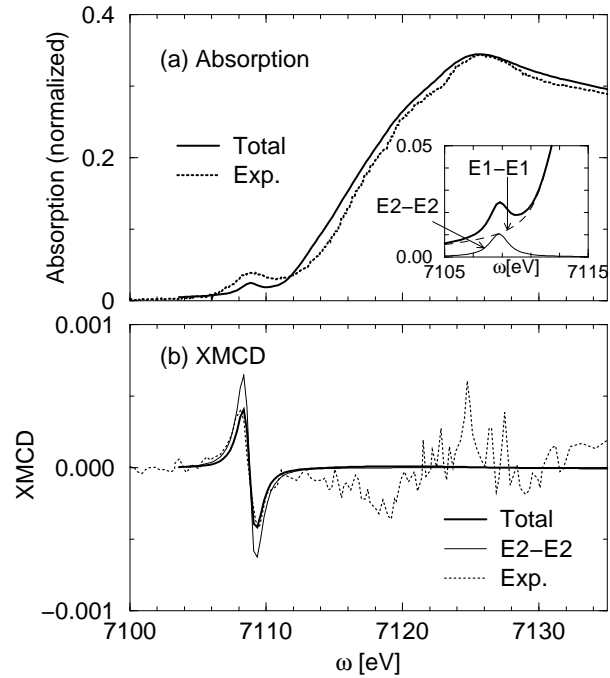}

\caption{Absorption coefficient (upper) and XMCD spectra (lower)
as a function of photon energy.
The solid and dotted lines represent the calculated and the experimental 
spectra,\cite{Matsumoto2000} respectively.
The inset is the decomposition of the total spectrum
into $A^{11}(\omega)$ and $A^{22}(\omega)$
in the pre-edge region.
\label{fig.absorption}}

\end{center}
\end{figure}

In the main $K$-edge region, the absorption coefficient is given by
\begin{equation}
 A^{11}_{\eta\eta}(\omega) = 2|M_1|^2\left(-\frac{1}{\pi}\right)
     {\rm Im}G_0(\omega),
\label{eq.A11_4p}
\end{equation}
where $G_0(\omega)$ is defined by eq.~(\ref{eq.g0}).
It is expressed by the sum over ${\bf k}$ and can be replaced by the integral 
of the $4p$ DOS. It is known in many transition-metal compounds that 
the absorption spectra are well reproduced in the wide range $20\sim 30$ eV 
by means of the $4p$ DOS given by the band calculation.\cite{Com2} 
In this paper, instead of carrying out the band calculation,
we assume the $4p$ DOS rising from the energy corresponding to $\omega=7111$ eV
with the band width as large as $30$ eV and sharp cutoff, so that it reproduces
the experimental absorption spectra in the main K-edge region 
(see Fig.~\ref{fig.absorption}).

Focusing on the contributions from the A sites in the pre-edge region, 
we have a more accurate form for $A^{11}_{\eta\eta}(\omega)$.
Equation (\ref{eq.A11_4p}) is modified by including the last term of 
eq.~(\ref{eq.g4p-4p}), 
\begin{equation}
 A^{11}_{\eta\eta'}(\omega) = |M_1|^2\sum_{\sigma}
  D^{11}_{\sigma\eta\beta_0,\sigma\eta'\beta_0}(\omega) , 
\label{eq.A11}
\end{equation}
with
\begin{equation}
 D^{11}_{\sigma\eta\beta_0,\sigma\eta'\beta_0}(\omega) = \frac{1}{2\pi i}
 \left\{ [G(\omega)]_{\sigma\eta'\beta_0,\sigma\eta\beta_0}^{*} -
 [G(\omega)]_{\sigma\eta\beta_0,\sigma\eta'\beta_0}\right\}.
\end{equation}
Here $G^{*}$ is a complex conjugate of $G$. 
Only the lowest energy state $|\beta_0\rangle$ 
in the $3d^5+3d^6\underline{L}$ configuration, which is equivalent to 
$|g\rangle$, is considered by the reason explained in the previous section.
The contribution of the last term of eq.~(\ref{eq.g4p-4p}) is, however,
one order of magnitude smaller than that from the first term,
and $A^{11}_{\eta\eta}(\omega)$ is practically determined by 
eq.~(\ref{eq.A11_4p}). Note that $A^{11}_{\eta\eta}(\omega)$ could include
the contribution of the $3d$ states at further neighbor iron sites 
if a larger cluster is considered.

The contributions of the $E$1-$E2$ process, $A^{12}_{\eta m}(\omega)$ 
and $A^{21}_{m\eta}(\omega)$, are canceled out after summing 
over the A sites, 
since they are proportional to $[G(\omega)]_{\sigma\eta\beta_0,\sigma\gamma}$ 
and $[G(\omega)]_{\sigma\gamma,\sigma\eta\beta_0}$ at each site, 
and their signs change between at the A$_1$ and A$_2$ sites. 
Therefore, the breaking of centrosymmetry could not influence
the absorption spectra.\cite{Com1}
The contribution of the $E$2-$E$2 process is given by 
\begin{equation}
 A^{22}_{mm'}(\omega) = |M_2|^2\sum_{\sigma\gamma\gamma'}
  \langle g|d_{m\sigma}|\gamma\rangle D^{22}_{\sigma\gamma,\sigma\gamma'}
  (\omega) \langle\gamma'|d_{m'\sigma}^{\dagger}|g\rangle ,
\end{equation}
with
\begin{equation}
 D^{22}_{\sigma\gamma,\sigma\gamma'}(\omega) = \frac{1}{2\pi i}
 \left\{ [G(\omega)]_{\sigma\gamma',\sigma\gamma}^{*} - 
 [G(\omega)]_{\sigma\gamma,\sigma\gamma'}\right\}.
\end{equation}

The upper panel in Fig.~\ref{fig.absorption} shows the calculated absorption
spectra in comparison with the experiment.\cite{Matsumoto2000}
Any reliable theoretical estimates of the core-level 
energy are not available. 
In addition, the $K$-edge energy is different for different experiments.\cite{Kanazawa2002,Kawata1995,Matsubara2005} 
Therefore, we have tentatively adjusted the energy separation between 
the $1s$-core level and the $4p$ states. 
Since $A^{22}_{mm}(\omega)$ is limited within the pre-edge region,
the spectra in the main $K$-edge region are dominated by 
$A^{11}_{\eta\eta}(\omega)$. The band bottom of the $4p$ DOS corresponds to
$\omega=7111$ eV. A tail in $A^{11}_{\eta\eta}(\omega)$ due to $\Gamma$ gives
a substantial contribution in the pre-edge region,as shown in the inset 
in the figure.
The total intensity in the pre-edge region is underestimated in comparison
with the experiment.\cite{Matsumoto2000}

When the x-ray is traveling along the direction opposite to the magnetization,
the absorption coefficient is different between the right-hand and left-hand 
circular polarizations. The XMCD is defined by the difference between them.
It is known that the XMCD is brought about by the SOI.
We neglect the SOI on the $4p$ states, since its
effect is expected to be very 
small in the pre-edge region. 
The lower panel in Fig.~\ref{fig.absorption} shows the calculated XMCD spectra 
in comparison with the experiment.\cite{Matsumoto2000} 
The calculated difference is divided by the value at the peak of the main edge 
in the absorption coefficient. Since no scale is shown for the XMCD
spectra in ref. \citen{Matsumoto2000}, the experimental 
curve is drawn in arbitrary scale. 
The $E$2-$E$2 process gives the largest contribution.  

Note that these results are obtained
for the A sites. For the B sites, the main $K$-edge spectra
are the same, but the pre-edge spectra could be different.
We need to consider the contribution from the B sites for quantitative comparison 
with the experiment.

\subsection{RXS spectra}

\begin{figure}[h]
\begin{center}
\includegraphics[width=8.0cm]{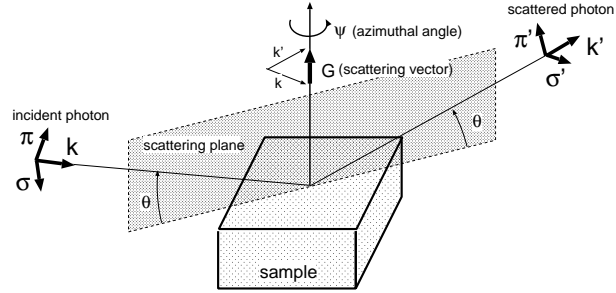}

\caption{Scattering geometry. Incident x-ray with momentum ${\bf k}$ 
and polarization $\sigma$ or $\pi$ is scattered into the state with 
momentum ${\bf k}'$ and polarization $\sigma'$ or $\pi'$.
\label{fig.geometry}}

\end{center}
\end{figure}

We consider the scattering geometry 
as illustrated in Fig.~\ref{fig.geometry}, where the incident x-ray 
with momentum ${\bf k}$, energy $\omega$, polarization ${\bf\epsilon}$
is scattered into the state with momentum ${\bf k}'$, energy $\omega$,
polarization ${\bf\epsilon'}$.
We define the scattering vector by ${\bf G}\equiv {\bf k}'-{\bf k}$.\cite{Com3}

By the same reason as the case of the absorption spectra,
the RXS amplitude may be given by a sum of amplitudes from each iron site. 
Then the scattering amplitude per unit cell is expressed as
\begin{eqnarray}
 F({\bf G},\omega) &=& r_0\Bigl{[} F^{\rm Th}({\bf G})
 {\bf \epsilon\cdot\epsilon'}
 + \sum_{\eta,\eta'}
   P_{\eta'}^{\mu'}F^{11}_{\eta'\eta}({\bf G},\omega) P_{\eta}^{\mu} 
 +\sum_{\eta,m}P_{\eta'}^{\mu'}F^{12}_{\eta'm}({\bf G},\omega) Q_{m}^{\mu} 
 \nonumber\\
 &+& \sum_{m',\eta}Q_{m'}^{\mu'}F^{21}_{m'\eta}({\bf G},\omega) 
  P_{\eta}^{\mu}
 + \sum_{m',m}Q_{m'}^{\mu'}F^{22}_{m'm}({\bf G},\omega) Q_{m}^{\mu}\Bigr{]},
\label{eq.rxs}
\end{eqnarray}
where the classical electron radius
$r_0\equiv e^2/(mc^2)=2.82\times 10^{-13}$ cm.
The first term represents Thomson scattering, which may be estimated as
\begin{equation}
 F^{\rm Th}({\bf G}) = \sum_{j} f^{0}_j({\bf G}) \exp(-i{\bf G}\cdot{\bf r}_j),
\end{equation}
where $f^{0}_j({\bf G})$ is the atomic form factor with $j$ running over 
not only iron sites but also oxygen sites. 

The remaining terms represent resonant scattering. They are defined by
\begin{equation}
 F^{\lambda'\lambda}_{\zeta',\zeta}({\bf G},\omega) = 
  \sum_{j}f^{\lambda'\lambda}_{j}(\omega)_{\zeta'\zeta}
  \exp(-i{\bf G}\cdot{\bf r}_j), \quad \lambda,\lambda'=1,2,
\end{equation}
where $f^{\lambda'\lambda}_{j}(\omega)_{\zeta'\zeta}$ is the resonant 
scattering amplitude at the $j$th iron site in the unit cell. 
For example, $f^{12}_{j}(\omega)$ is defined by
\begin{equation}
 f^{12}_{j}(\omega)_{\eta m} = mc^2 \sum_{n\sigma}
 \frac{\langle g| T^{\rm E1\dagger}_{\eta\sigma}(j)|n\rangle
 \langle n | T^{\rm E2}_{m\sigma}(j)|g\rangle}
 {\omega-E_n+E_g + i\Gamma}.
\end{equation}
Other components are similarly defined.
In eq.~(\ref{eq.rxs}), $P^{\mu}_{\eta}$ and $Q^{\mu}_{m}$
represent the geometrical factors of the E1 and E2 transitions for 
the incident x-ray with polarization $\mu=\sigma$ or $\pi$,
while $P^{\mu'}_{\eta}$ and $Q^{\mu'}_{m}$ represent those 
for the scattered x-ray with polarization $\mu'=\sigma'$ or $\pi'$.
Their general expressions are summarized in Appendix.

The resonant terms at the A sites are expressed by using the resolvent given 
in Sec. III:
\begin{eqnarray}
 f^{11}_{\rm A}(\omega)_{\eta'\eta} &=& mc^2 |M_1|^2 \sum_{\sigma}
  [G(\omega)]_{\sigma\eta',\sigma\eta}, \label{eq.f11}\\
 f^{12}_{\rm A}(\omega)_{\eta'm} &=& mc^2 M^{*}_1 M_2\sum_{\sigma\gamma}
  [G(\omega)]_{\sigma\eta'\beta_0,\sigma\gamma}
  \langle\gamma|d^{\dagger}_{m\sigma}|g\rangle, \label{eq.f12}\\
 f^{21}_{\rm A}(\omega)_{m'\eta} &=& mc^2 M_2^{*} M_1\sum_{\sigma\gamma}
  \langle g|d_{m'\sigma}|\gamma\rangle
  [G(\omega)]_{\sigma\gamma,\sigma\eta\beta_0}, \label{eq.f21}\\
 f^{22}_{\rm A}(\omega)_{m'm} &=& mc^2 |M_2|^{2} \sum_{\sigma\gamma\gamma'}
  \langle g|d_{m'\sigma}|\gamma\rangle
  [G(\omega)]_{\sigma\gamma,\sigma\gamma'}
  \langle\gamma'|d_{m\sigma}|g\rangle .
\end{eqnarray}
Amplitudes $f^{12}_{\rm A}(\omega)_{\eta'm}$ and 
$f^{21}_{\rm A}(\omega)_{m'\eta}$ change their signs between 
the A$_1$ and
A$_2$ sites, respectively. They depend also on the direction of the local 
magnetic moment.  A careful examination of eqs.~(\ref{eq.f12}) and 
(\ref{eq.f21}) leads us to the expression,
\begin{eqnarray}
f^{12}_{\rm A}(\omega) &=&
\begin{array}{cccccc}
& x^2-y^2 & 3z^2-r^2 & yz & zx & xy \\
\begin{array}{c}
x \\
y \\
z \\
\end{array} & \left(
\begin{array}{c}
b(\omega)n_x \\
b(\omega)n_y \\
-2b(\omega)n_z \\
\end{array} \right. &
\begin{array}{c}
c(\omega)n_x \\
-c(\omega)n_y \\
0 \\
\end{array} &
\begin{array}{c}
a(\omega) \\
d(\omega)n_z \\
-d(\omega)n_y \\
\end{array} &
\begin{array}{c}
-d(\omega)n_z \\
a(\omega) \\
d(\omega)n_x \\
\end{array} &
\left.
\begin{array}{c}
d(\omega)n_y \\
-d(\omega)n_x \\
a(\omega) \\
\end{array}
\right) , \\
\end{array}
\label{eq.F12}\\
f^{21}_{\rm A}(\omega) &=&
\begin{array}{cccc}
& x & y & z \\
\begin{array}{c}
x^2-y^2 \\
3z^2-r^2 \\
yz \\
zx \\
xy \\
\end{array} &
\left(
\begin{array}{c}
b(\omega)n_x \\
c(\omega)n_x \\
-a(\omega) \\
-d(\omega)n_z \\
d(\omega)n_y \\
\end{array}
\right. &
\begin{array}{c}
b(\omega)n_y \\
-c(\omega)n_y \\
d(\omega)n_z \\
-a(\omega) \\
-d(\omega)n_x \\
\end{array} &
\left.
\begin{array}{c}
-2b(\omega)n_z \\
0 \\
-d(\omega)n_y \\
d(\omega)n_x \\
-a(\omega) \\
\end{array}
\right) , \\
\end{array}
\label{eq.F21}
\end{eqnarray}
where $(n_x,n_y,n_z)$ represents the direction cosine of the local magnetic
moment vector centered at each iron atom.
Note that the local magnetic moment at the A sites is opposite to the total
magnetization.
The component $a(\omega)$, which is independent of the direction of 
the local magnetic moment, exists even in the absence of the SOI.
On the other hand, $b(\omega)$, $c(\omega)$, and $d(\omega)$,
which are one order of magnitude smaller than $a(\omega)$,
disappear without the SOI. 
All these components are appreciable only in a narrow pre-edge region.

For the resonant terms at the B sites, $f^{11}_{\rm B}(\omega)$ may be expressed by 
\begin{equation}
 f^{11}_{\rm B}(\omega)_{\eta'\eta} = mc^2 |M_1|^2 
  2G_0(\omega)\delta_{\eta'\eta},
\label{eq.f11B}
\end{equation}
where the contribution like the last term of eq.~(\ref{eq.g4p-4p}) 
does not exist. 
The $f^{12}_{\rm B}(\omega)$ and $f^{21}_{\rm B}(\omega)$ disappear 
because of centrosymmetry. 
We expect $|f^{22}_{\rm B}(\omega)|<<|f^{11}_{\rm B}(\omega)|$ in the pre-edge
region. This contrast with the absorption coefficient, where the $E$2-$E$2
contribution is comparable to the $E$1-$E$1 contribution.
The reason is that the scattering amplitude is affected 
by $G_0(\omega)$ itself, whose real part is about two orders of magnitude 
larger than the imaginary part, while only the imaginary part contributes
to the absorption coefficient. Note that, though it is small, 
$f^{22}_{\rm B}(\omega)$ could give rise to the magnetic scattering amplitude, 
which study is outside of the purpose of this paper.

In the following, we analyze the RXS spectra at several Bragg spots, 
focusing on the $\sigma-\sigma'$ channel.

\subsubsection{${\bf G}=(002)$ and $(006)$}

For position vectors given by eq.~(\ref{eq.position}), 
the phase factors $\exp({-i{\bf G\cdot r}_j})$ are $1$ at the
A$_1$ sites, $-1$ 
at the A$_2$ sites, 
and $\mp i,\mp i,\mp i,\mp i,\pm i,\pm i,\pm i,\pm i,\pm i,\pm i,\pm i,\pm i,
\mp i,\mp i,\mp i,\mp i,\mp i$ (upper and lower signs correspond to
$(002)$ and $(006)$, respectively) at the B sites.
Thomson scattering amplitude as well as all the resonant terms
are canceled out except for $f^{12}_{\rm A}(\omega)$ and 
$f^{21}_{\rm A}(\omega)$, due to the phase factors.
Therefore, these Bragg spots are suitable to investigate noncentrosymmetric
effects on the RXS.
Several experiments of RXS have actually been carried out on these spots,
\cite{Garcia2001,Kanazawa2002} but the dependence of the spectra
on the magnetization direction has not been studied yet.
We calculate the RXS spectra and analyze such dependence.

We consider the situation that the scattering plane contains a vector 
$(1,-1,0)$ (see Fig.~\ref{fig.geometry}).
The geometrical factors $P^{\sigma}$, $P^{\sigma'}$, $Q^{\sigma}$, and
$Q^{\sigma'}$ are given by putting $\psi=\pi/4$ in the expressions
in Appendix.
We assume that the local magnetic moment on the A sites is along 
the ${\bf n}=(n_x,n_y,0)$ direction.
Then, using eqs.~(\ref{eq.F12}) and (\ref{eq.F21}),
we have the scattering amplitude in the $\sigma-\sigma'$ channel, 
\begin{eqnarray}
 F({\bf G},\omega) &=& r_0
 \sum_{\eta,m}\Bigl[P^{\sigma'}_{\eta}F^{12}_{\eta m}({\bf G},\omega)
 Q^{\sigma}_{m} + Q^{\sigma'}_{m}F^{21}_{m\eta}({\bf G},\omega)
 P^{\sigma}_{\eta}\Bigl] \nonumber \\
 &=& 8r_0\bigl[2\sin\theta a(\omega) + \sqrt{2}(n_x-n_y)\cos\theta b(\omega)
 \bigr],
\end{eqnarray}
where Bragg angle $\theta$ is $12.0$ and $38.5$ degrees for $(002)$ and 
$(006)$, respectively.
Let $I_{+}({\bf G},\omega)$ and $I_{-}({\bf G},\omega)$ be the intensities 
per unit cell for the direction of the magnetic moment ${\bf n}$ and 
the reverse, respectively. Then the average and the difference of the 
intensities are given by
\begin{eqnarray}
 I^{\sigma-\sigma'}({\bf G},\omega) &\equiv& \frac{1}{2}\left(
 I_{+}^{\sigma-\sigma'}({\bf G},\omega) 
   +I_{-}^{\sigma-\sigma'}({\bf G},\omega)\right) \nonumber\\
   &  =   & 256r_0^2\{\sin^{2}\theta|a(\omega)|^2
                   +\frac{1}{2}\cos^{2}\theta(n_x-n_y)^2|b(\omega)|^2 \},\\
\Delta I^{\sigma-\sigma'}({\bf G},\omega) &\equiv& 
 I_{+}^{\sigma-\sigma'}({\bf G},\omega)-I_{-}^{\sigma-\sigma'}({\bf G},\omega) 
 \nonumber\\
   &  =   & 128\sqrt{2}r_0^2\sin 2\theta(n_x-n_y)[a(\omega)^{*}b(\omega)
                     +{\rm c.c.}].
\end{eqnarray}
Since $b(\omega)$ is one order of magnitude smaller than $a(\omega)$, 
the average intensity $I^{\sigma-\sigma'}(\omega)$ is dominated by 
$|a(\omega)|^2$. On the other hand,
the difference spectra arise from the interference between the terms of
$a(\omega)$ and $b(\omega)$.

\begin{figure}[h]
\begin{center}
\includegraphics[width=8.0cm]{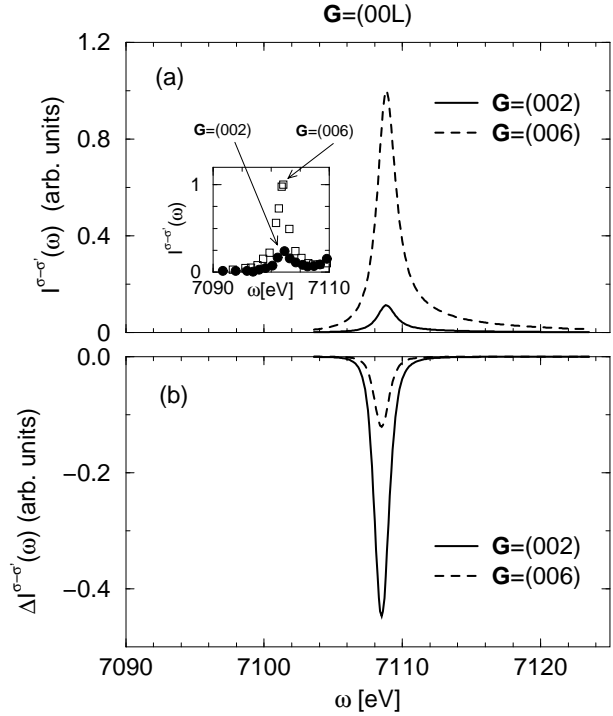}

\caption{Scattering intensity as a function of photon energy
in the pre-edge region in the $\sigma-\sigma'$ channel.
Panel (a) shows the average intensity. Panel (b) shows the intensity 
difference when the magnetic field is applied along
the $[-1,1,0]$ direction and the reverse, which is
divided by the peak value of the average intensity.
The solid and broken lines represent the intensities for ${\bf G}=(002)$ 
and $(006)$, respectively. The inset on panel (a) represents the experimental
curves taken from ref.~\citen{Kanazawa2002}.
\label{fig.G00L}}

\end{center}
\end{figure}

Figure \ref{fig.G00L} shows $I^{\sigma-\sigma'}({\bf G},\omega)$
and $\Delta I^{\sigma-\sigma'}({\bf G},\omega)$ calculated with 
${\bf n}=(1/\sqrt{2},-1/\sqrt{2},0)$.
This magnetization direction corresponds to the magnetic field applied 
along the $[-1,1,0]$ direction.  The $I^{\sigma-\sigma'}({\bf G},\omega)$ 
is concentrated in a narrow pre-edge region, and becomes larger for 
${\bf G}=(006)$,
in consistent with the experiments.\cite{Garcia2001,Kanazawa2002}
The $\Delta I^{\sigma-\sigma'}({\bf G},\omega)$ is relatively large,
only one order of magnitude smaller than the average intensity.
This intensity difference is a consequence of breaking 
both the local centrosymmetry and the time-reversal symmetry.
It would not be hard to detect such a difference.

\subsubsection{${\bf G}=(222)$}

For position vectors given by eq.~(\ref{eq.position}),
the phase factors $\exp({-i{\bf G\cdot r}_j})$ 
are $1$ at the A$_1$ sites, $-1$ at the A$_2$ sites, 
and $i$ at all B sites.
Therefore this spot is not prohibited.

Thomson scattering gives the largest contribution;
the contribution from the A sites are canceled out, but those from the B sites 
and oxygen sites remain, resulting in 
\begin{equation}
 F^{\rm Th}({\bf G}) = 16if_{\rm B} - 32if_{\rm O},
\end{equation}
where $f_{\rm B}=(1/2)[f({\rm Fe}^{2+})+f({\rm Fe}^{3+})]$ is the form 
factor of iron at the B sites and $f_{\rm O}=f({\rm O}^{2-})$ is that of oxygen. 
They are evaluated from the atomic values tabulated 
in ref.~\citen{Coppens}. 
As regards the resonant terms, $f^{11}_{\rm A}(\omega)$ 
and $f^{22}_{\rm A}(\omega)$ are canceled out at the A sites
due to the phase factor, but the contribution from the B sites remains.
We have
\begin{equation}
 [F^{11}({\bf G},\omega)]_{\eta'\eta} \equiv F^{11}_{0}({\bf G},\omega)
 \delta_{\eta'\eta} = 16if^{11}_{\rm B}(\omega) \delta_{\eta'\eta},
\end{equation}
where $f^{11}_{\rm B}(\omega)$ is evaluated from eq.~(\ref{eq.f11B}).
Note that the atomic calculation of $f^{11}_{\rm B}(\omega)$ 
would contain large errors in the pre-edge region, 
because a single level of $4p$ states in an atom changes into an energy band 
with width as large as $\sim 20$ eV in solids. We list the calculated values in 
Table \ref{table.2}.
The Thomson scattering amplitude is much larger than the resonant term.
These values are much larger than those reported in 
ref.~\citen{Matsubara2005}.
Thus, the average intensity is given by
\begin{equation}
 I^{\sigma-\sigma'}({\bf G},\omega) 
    = r_0^2|F^{\rm Th}({\bf G})+F^{11}_{0}({\bf G},\omega)|^2. 
\end{equation}

\begin{table}[tb]
\caption{$F^{\rm Th}({\bf G})$ and $F^{11}({\bf G},\omega_0)$ with $\omega_0$ 
being the pre-edge absorption peak.}
\label{table.2}
\begin{tabular}{ccc}
\hline
$hkl$   & $F^{\rm Th}({\bf G})$ & $[F^{11}({\bf G},\omega_0)]_{\eta\eta}$ \\
$(222)$ & $i\,123.1$       & $0.68-i\,10.24$  \\
$(333)$ & $148.7-i\,148.7$ & $-32.52+i\,5.78$  \\
\hline
\end{tabular}
\end{table}

Next we analyze the dependence on the direction of applied magnetic 
field in accordance with the experiment.\cite{Matsubara2005} 
The scattering plane is set to contain a vector $(1,-1,0)$ 
with applying magnetic field along the $[11\overline{2}]$ direction 
and the reverse.
The geometrical factors $P^{\sigma}$, $P^{\sigma'}$, $Q^{\sigma}$, and
$Q^{\sigma'}$ are given by putting $\psi=\pi/4$ in the expressions in Appendix. 
Substituting $(\mp 1/\sqrt{6},\mp 1/\sqrt{6},\pm 2/\sqrt{6})$ 
for $(n_x,n_y,n_z)$ in eqs.~(\ref{eq.F12}) and (\ref{eq.F21}),
we obtain the scattering amplitude in the $\sigma-\sigma'$ channel,
\begin{equation}
 \sum_{\eta,m}\Bigl[P^{\sigma'}_{\eta}[F^{12}({\bf G},\omega)]_{\eta m}
 Q^{\sigma}_{m} + Q^{\sigma'}_{m}[F^{21}({\bf G},\omega)]_{m\eta}
 P^{\sigma}_{\eta}\Bigl]
 = 8\cos\theta [\delta_0 a(\omega) \pm \delta_1 c(\omega) 
 \pm \delta_2 d(\omega)],
\end{equation}
where the upper (lower) signs correspond to upper (lower) signs of 
$(n_x,n_y,n_z)$, and $\delta_0= -(2/3)(1/\sqrt{2}+2/\sqrt{3})$,
                     $\delta_1= -\sqrt{2}/3$,
                     $\delta_2= (2/3)(1-\sqrt{2/3})$.
Bragg angle $\theta$ is $21.1$ degrees.
Since the direction of the local magnetic moment on the 
A sites is 
opposite to the direction of the applied magnetic field, 
we define the intensity difference as the value with the upper sign for
${\bf n}$ minus the value with the lower sign for ${\bf n}$.
As a consequence, we have
\begin{equation}
\Delta I^{\sigma-\sigma'}(\omega) 
    =  16r_0^2\cos\theta\{ [F^{\rm Th}({\bf G})^{*}+F^{11}_{0}(\omega)^{*}]
         [\delta_1 c(\omega) + \delta_2 d(\omega)] + {\rm c.c.}\}.
\label{eq.diff.222}
\end{equation}
The intensity difference arises from the interference between the term of
$F^{\rm Th}({\bf G})+F^{11}_{0}({\bf G},\omega)$ and the terms of $c(\omega)$ 
and $d(\omega)$.

\begin{figure}[h]
\begin{center}
\includegraphics[width=8.0cm]{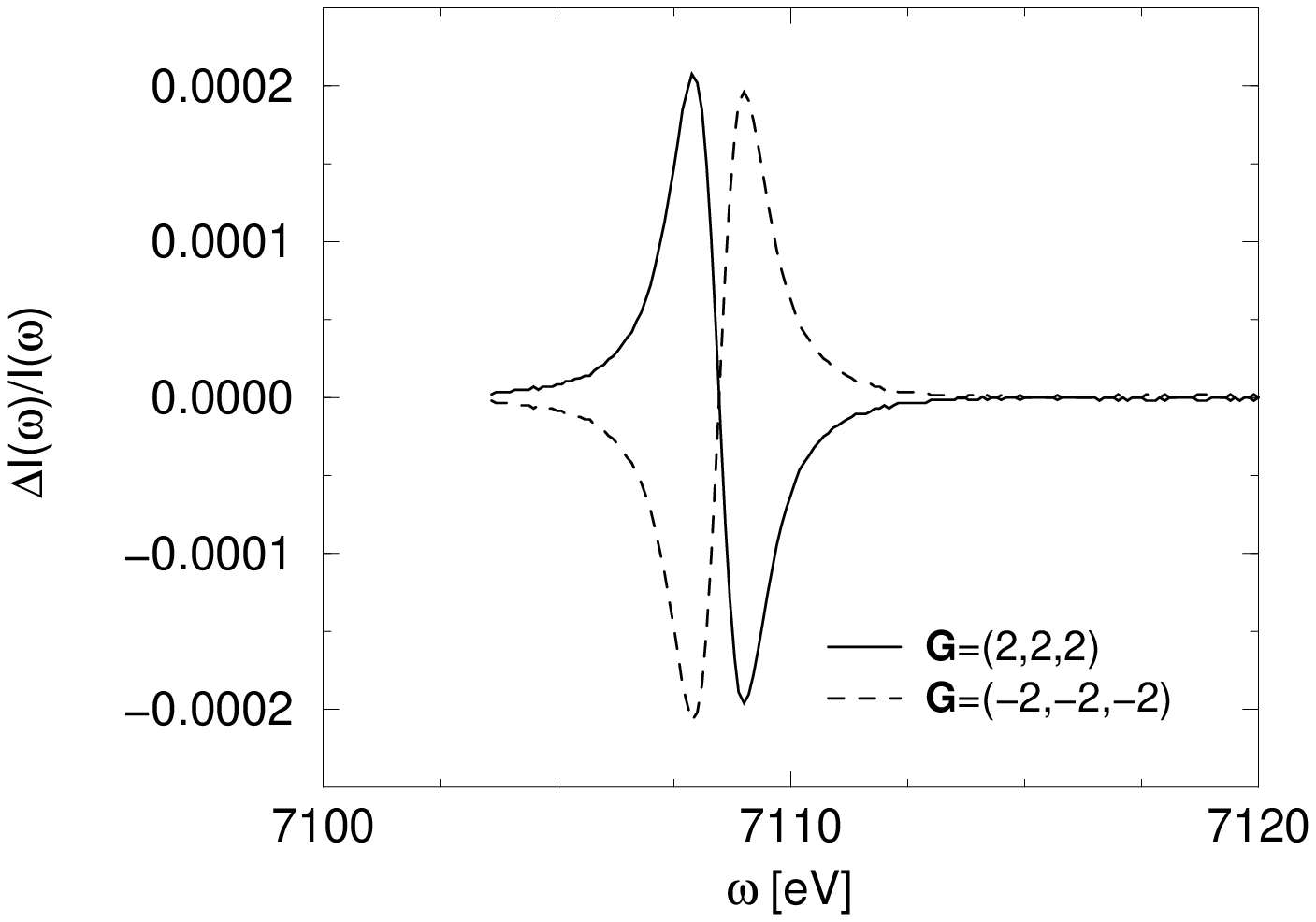}
\caption{Relative intensity difference 
$\Delta I^{\sigma-\sigma'}({\bf G},\omega)/
I^{\sigma-\sigma'}({\bf G},\omega)$ as a function of photon energy
at ${\bf G}=(222)$ and 
${\bf G}=(\overline{2}\overline{2}\overline{2})$ in the pre-edge region.
\label{fig.G222}
}
\end{center}
\end{figure}

Figure \ref{fig.G222} shows the relative intensity difference
$\Delta I^{\sigma-\sigma'}({\bf G},\omega)/I^{\sigma-\sigma'}({\bf G},\omega)$
in the pre-edge region. The spectral shape takes a peculiar "dispersion" form.
This may be explained as follows. 
The factor $F^{\rm Th}({\bf G})+F^{11}({\bf G},\omega)$ 
at $(222)$ spot is very close 
to a pure imaginary number as shown in Table \ref{table.2}. Another factor
$c(\omega)$ or $d(\omega)$ is given by a resolvent matrix element multiplied by
$M^{*}_1 M_2$ which is a pure imaginary number.
Thereby the product of the two factors in eq.~(\ref{eq.diff.222})
becomes a resolvent matrix element multiplied by a real number. 
By adding its complex conjugate,
eq.~(\ref{eq.diff.222}) becomes proportional to the real part of the resolvent
matrix element. The real part of the Green function usually takes a 
"dispersion" form as a function of energy.

The intensity difference in the experiment by Matsubara 
\textit{et al}.\cite{Matsubara2005} 
is different from the one calculated here. 
It extends over a region much 
wider than the region of the pre-edge absorption spectra with an order of
magnitude larger intensity, which behavior is quite unusual. 
Also the shape is different from a ``dispersion" form. 
Matsubara \textit{et al}. claimed that the spectra they found arise from a 
``magnetoelectric" amplitude, that is, a consequence of breaking
both the centrosymmetry and the time-reversal symmetry.
According to the present analysis, this claim has no ground.
In this connection, we would like to draw attention to the similar RXS 
experiment for MnCr$_2$O$_4$, where the shape and strength quite similar 
to the curve calculated above have been observed
at the Mn pre-K-edge.\cite{Thesis} 
In this material, Mn atoms occupy at the A sites in spinel
structure. Since the Mn pre-edge spectrum selects only the A site contribution,
this experimental result indicates that the calculated spectra correspond
to the ``magnetoelectric" signal.
Note that a similar ``dispersion" form of the spectra 
has been observed\cite{Arima2005}
and theoretically analyzed\cite{DiMatteo2006,Lovesey2007-2}
at the Fe $K$ edge of GaFeO$_3$.

Finally we comment on what happens on the intensity difference 
when the scattering vector is reversed. Different from the conventional case,
the signal is reversed with keeping the shape, as shown in 
Fig.~\ref{fig.G222}. The $F^{12}({\bf G},\omega)$ and $F^{12}({\bf G},\omega)$ 
are unaltered because the phase factors at the 
A sites are the same 
with reversing ${\bf G}$. On the other hand, 
$F^{\rm Th}({\bf G})+F^{11}({\bf G},\omega)$ changes its sign, 
because the phase factors at the 
B sites is changed from $i$ to $-i$, 
resulting in the sign change in eq.~(\ref{eq.diff.222}).

\subsection{${\bf G}=(333)$}

The phase factors $\exp({-i{\bf G\cdot r}_j})$ are 1 at the
A$_1$ sites, 
$-i$ at the A$_2$ sites, 
$\exp(i3\pi/4)$, $\exp(i3\pi/4)$, $\exp(i3\pi/4)$, $\exp(i3\pi/4)$, 
$\exp(-i\pi/4)$, $\exp(-i\pi/4)$, $\exp(-i\pi/4)$, $\exp(-i\pi/4)$, 
$\exp(-i\pi/4)$, $\exp(-i\pi/4)$, $\exp(-i\pi/4)$, $\exp(-i\pi/4)$, 
$\exp(i\pi/4)$, $\exp(i\pi/4)$, $\exp(i\pi/4)$, $\exp(i\pi/4)$
at the B sites, respectively.
The
Thomson scattering amplitude and the resonant term 
$F^{11}({\bf G},\omega)$ are given by
\begin{eqnarray}
 F^{\rm Th}({\bf G}) &=& 4(1-i)(f_{\rm A}+\sqrt{2}f_{\rm B}), \\
 F^{11}({\bf G},\omega)_{\eta'\eta} &\equiv& F^{11}_{0}({\bf G},\omega)
 \delta_{\eta'\eta} = 4(1-i)[f^{11}_{\rm A}(\omega) 
   + \sqrt{2}f^{11}_{B}(\omega)]\delta_{\eta'\eta}.
\end{eqnarray}
These amplitudes are evaluated by using atomic form factors for $f_{\rm A}$
and $f_{\rm B}$ and eq.~(\ref{eq.f11B}) for $f^{11}_{\rm A}(\omega)$
and $f^{11}_{\rm B}(\omega)$. The results  which are listed in 
Table \ref{table.2}.
In addition, we have the E$_1$-E$_2$ term,
\begin{equation}
 \sum_{\eta,m}\Bigl[P^{\sigma'}_{\eta}[F^{12}({\bf G},\omega)]_{\eta m}
 Q^{\sigma}_{m} + Q^{\sigma'}_{m}[F^{21}({\bf G},\omega)]_{m\eta}
 P^{\sigma}_{\eta}\Bigl]
 = 4(1+i)\cos\theta [\delta_0 a(\omega) \pm \delta_1 c(\omega) 
 \pm \delta_2 d(\omega)],
\end{equation}
with Bragg angle $\theta=32.6$ degrees. As a result, we obtain the intensity
difference as
\begin{equation}
\Delta I^{\sigma-\sigma'}(\omega) 
    =  8r_0^2\cos\theta\{ [F^{\rm Th}({\bf G})^{*}+F^{11}_{0}(\omega)^{*}]
     (1+i)[\delta_1 c(\omega) + \delta_2 d(\omega)] + {\rm c.c.}\}.
\label{eq.diff.333}
\end{equation}
Since $F^{\rm Th}({\bf G})^{*}$ is proportional to $(1+i)$, the right hand
side of eq.~(\ref{eq.diff.333}) is nearly proportional to the real part of
a resolvent matrix element. Therefore we would expect a ``dispersion" form
of spectra as a function of photon energy.

\begin{figure}[h]
\begin{center}
\includegraphics[width=8.0cm]{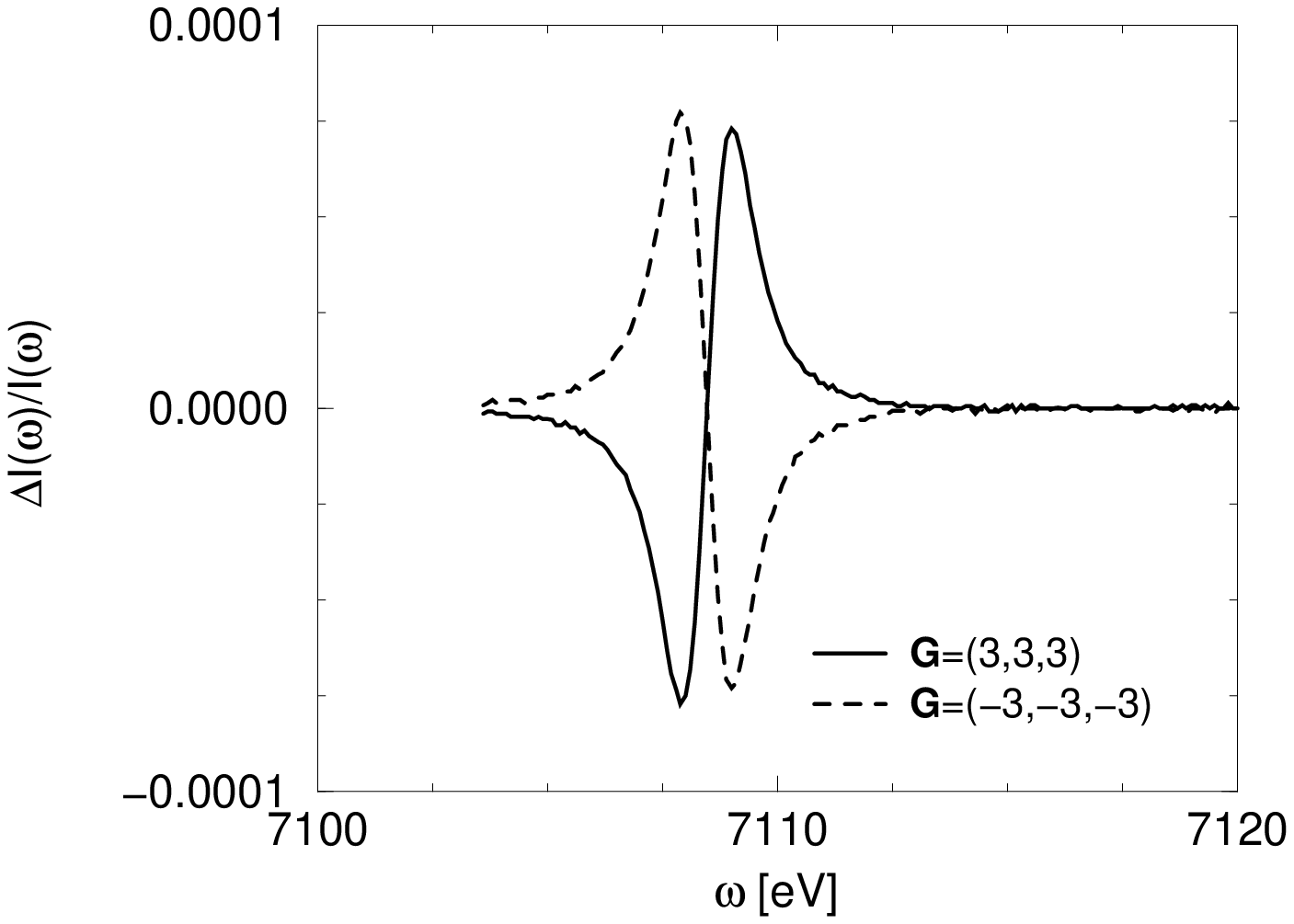}
\caption{Relative intensity difference 
$\Delta I^{\sigma-\sigma'}({\bf G},\omega)/
I^{\sigma-\sigma'}({\bf G},\omega)$
as a function of photon energy at
${\bf G}=(333)$ and ${\bf G}=(\overline{3}\overline{3}\overline{3})$
in the pre-edge region.
\label{fig.G333}
}
\end{center}
\end{figure}

Figure \ref{fig.G333} shows the relative intensity difference 
$\Delta I^{\sigma-\sigma'}({\bf G},\omega)/I^{\sigma-\sigma'}({\bf G},\omega)$
thus evaluated. The calculated value has
the same size of magnitude as 
the experimental one\cite{Matsubara2005}
and is nearly half of the calculated one for 
$(222)$.  Note that the experimental value at $(222)$ 
is about two order of magnitude larger than the values at $(333)$.

\subsection{${\bf G}=(444)$}

The phase factors $\exp({-i{\bf G\cdot r}_j})$ are $1$ at the
A sites and $-1$ at the B sites. 
Therefore, the $E$1-$E$2 terms $F^{12}({\bf G},\omega)$ and
$F^{21}({\bf G},\omega)$ vanish due to the cancellation 
between the A$_1$ and
A$_2$ sites. In the experiment,\cite{Matsubara2005}
the intensity dependence is found negligible.
If we take seriously this fact, it means that the contribution of 
the magnetic scattering amplitude $F^{22}({\bf G},\omega)$
is quite small.

\section{Concluding Remarks}

We have studied how the breaking of centrosymmetry affects the RXS spectra 
through a microscopic calculation for magnetite.
The centrosymmetry is locally broken at tetrahedral (A) sites.
In such a circumstance, the $4p$ states strongly hybridize with the $3d$ 
states through neighboring oxygen $2p$ states, giving rise to 
the non-vanishing contribution of the $E$1-$E$2 process in the RXS spectra. 
This observation is substantiated by introducing a microscopic model
of a FeO$_4$ cluster with the $4p$ states forming a band and the $3d$ states 
forming multiplet structures. We have calculated the RXS spectra 
with the help of the resolvent formalism.
It is shown that the hybridization changes its sign between 
the A$_1$
and A$_2$ sites and accordingly the local amplitude from the $E$1-$E$2 
process changes its sign. This sign change causes non-vanishing RXS intensities
at the forbidden spots $(002)$ and $(006)$.
The spectra are concentrated in a narrow pre-edge region  
with intensities larger at $(006)$ than at $(002)$, 
in agreement with the experiment.
In addition, we have carefully analyzed the scattering matrix for the 
$E$1-$E$2 process, which depends on the direction of the applied magnetic 
field. Such dependence is only possible when both centrosymmetry and
time-reversal symmetry are broken. Through this analysis, we have obtained 
large dependences of intensity at $(002)$ and $(006)$ spots. 
We hope that this dependence could be observed in future experiments.

We have also analyzed the dependence on the direction of the applied magnetic 
field at $(222)$, $(333)$ and $(444)$ spots in connection 
with the experiment.
These spots are allowed with large Thomson scattering amplitudes.
Having calculated the intensities for two opposite directions
of the applied magnetic field, we have obtained their difference
with the same order of magnitude at both $(222)$ and $(333)$ spots 
but negligible difference at $(444)$.
The intensity difference is found to has a ``dispersion" form as a function
of photon energy, which is concentrated in a narrow pre-edge region. 
In the experiment by Matsubara \textit{et al}.,\cite{Matsubara2005} however,
the intensity difference at the $(222)$ spot is distributed over the region
much wider than the pre-edge region with the spectral shape quite 
different from the ``dispersion" form. 
The observed intensity difference at $(222)$ is two orders of 
magnitude larger than the one
at $(333)$. These behaviors seem unusual and hard to explain.
Matsubara \textit{et al}. claimed that the difference at the
$(222)$ spot arises from 
a ``magnetoelectric" amplitude, that is, a consequence of breaking 
both centrosymmetry and time-reversal symmetry. This claim has no ground,
according to the analysis in this paper. A ``dispersion" form of spectral
shape has been observed in the experiment at the Mn pre-$K$-edge in 
MnCr$_2$O$_4$.\cite{Thesis}
Since Mn atoms are occupying at the 
A sites in spinel structure, the pre-$K$-edge
signal selects the contribution from the A sites. This experiment
suggests that the calculated spectra correspond to the ``magnetoelectric"
signal in the magnetite. The large signal at the $(222)$ spot might be 
related to B sites.
We hope experiments in future clarify the situation in magnetite.

\begin{acknowledgments}

We would like to thank Dr. Matsubara for providing us with his thesis and
for valuable discussions. This work was partly supported by Grant-in-Aid 
for Scientific Research from the Ministry of Education, Culture, Sport, 
Science, and Technology, Japan.

\end{acknowledgments}

\appendix

\section{Geometrical factors}

We briefly describe a derivation of geometrical factors.
A geometrical setting of x-ray scattering adopted in the
present work is shown in Fig. \ref{fig.geometry}.
We introduce three coordinate systems,
$(x',y',z')$, $(x'',y'',z'')$, and $(x''',y''',z''')$.
The first coordinate system is attached to the incident
(scattered) photon
with its $z'$ axis being parallel to
$\textbf{k} (\textbf{k}')$. Its $x'$ and $y'$ axes are
perpendicular and parallel to the scattering plane,
respectively.
The second coordinate system is used as the definition of
the origin of the azimuthal angle $\psi$.
Its $z''$ axis is aligned to $\textbf{G}$ direction
and $y''$ axis is in the scattering plane at $\psi=0$
with $x''=x'$.
The third coordinate system is fixed
to the crystal.

These three coordinate systems can be connected by the
Euler rotation with the choices of
appropriate Euler angles.
Here we use the same definition of the Euler rotation
adopted in Rose's book.\cite{Rose1957}
From $(x'',y'',z'')$ to $(x',y',z')$ coordinate
systems, the Euler angles are chosen as
$\left(\frac{\pi}{2},\frac{\pi}{2}\pm\theta,-\frac{\pi}{2}
\right)$
where $\theta$ represents the Bragg angle.
The upper (lower) sign is for the incident (scattered)
photon. Hereafter, we restrict our discussion on the
incident photon case
alone, since the results for the scattered photon are
obtained by replacing every $\theta$ with $-\theta$.
The Euler angles of the rotation from the $(x'',y'',z'')$
coordinate system to the $(x''',y''',z''')$ coordinate
system are given by $(\alpha,\beta,0)$ where
$\alpha$ and $\beta$ are the azimuthal and the
polar angles of $\textbf{G}$, respectively.

In order to calculate the geometrical factors, we
start with writing down the basis
corresponding to the $\sigma$ and $\pi$ polarizations.
For dipole transition, it is simple since
$\mbox{\boldmath$\epsilon$}^{\sigma} = \textbf{e}_{x'}$ and
$\mbox{\boldmath$\epsilon$}^{\pi} = -\textbf{e}_{y'}$ hold
in the present setting. Here $\textbf{e}_j$ denotes
the unit vector directed to $j$ axis. Then,
the geometrical factors $\{ P_{j}^{\mu} \}$ ($\mu=\sigma, \pi$)
are defined by the following relations.
\begin{eqnarray}
\mbox{\boldmath$\epsilon$}^{\sigma} \cdot {\bf r}
& = & x'= P_{x}^{\sigma} x''' + P_{y}^{\sigma} y'''
+ P_{z}^{\sigma} z''', \label{eq.defPs} \\
\mbox{\boldmath$\epsilon$}^{\pi} \cdot {\bf r}
& = & -y' =
P_{x}^{\pi} x''' + P_{y}^{\pi} y''' + P_{z}^{\pi} z''',
\label{eq.defPp}
\end{eqnarray}
where $\textbf{r}$ is an arbitrary position vector.
Similarly, the geometrical factors $\{Q_{n}^{\mu} \}$
are defined by the quantity
$(\textbf{k} \cdot \textbf{r})
(\mbox{\boldmath$\epsilon$} \cdot \textbf{r})$
appearing in the multipole expansion of the scattering
amplitude. Here $n=1,2,3,4$, and $5$ correspond to
the quadrupole basis $x^2-y^2, 3z^2-r^2, yz, zx$,
and $xy$, respectively.
By noticing the fact that $\textbf{k} =|\textbf{k}|
\textbf{e}_{z'}$, we define
$\{Q_{n}^{\mu} \}$ in the following relations.
\begin{eqnarray}
(\textbf{k} \cdot \textbf{r})
(\mbox{\boldmath$\epsilon$}^{\sigma} \cdot \textbf{r})
&\propto& z_4' = \sum_{n=1}^{5} Q_{n}^{\sigma} z_n ''',
\label{eq.defQs} \\
(\textbf{k} \cdot \textbf{r})
(\mbox{\boldmath$\epsilon$}^{\pi} \cdot \textbf{r})
&\propto& -z_3' = \sum_{n=1}^{5} Q_{n}^{\sigma} z_n ''',
\label{eq.defQp}
\end{eqnarray}
where $z_1=\frac{\sqrt{3}}{2}(x^2-y^2)$
$z_2=\frac{1}{2}(3z^2-r^2)$, $z_3=\sqrt{3} yz$,
$z_4=\sqrt{3} zx$, and $z_5=\sqrt{3} xy$.

From eqs. (\ref{eq.defPs}) $\sim$ (\ref{eq.defQp}), the geometrical factors 
are evaluated by expressing $x', y'$ in terms of  $x''', y''', z'''$ 
for the dipole transition and $z'_{4}, z'_{3}$ in terms of  $z'''_{\mu}$ 
for the quadrupole transition, respectively, with the help of rotation matrix.
The final results for the $\sigma$ polarization with the incident photon
are as follows:
\begin{eqnarray}
P_{x}^{\sigma} & = &
\cos \alpha \cos \beta \cos \psi + \sin \alpha \sin \psi, \\
P_{y}^{\sigma} & = &
\sin \alpha \cos \beta \cos \psi - \cos \alpha \sin \psi, \\
P_{z}^{\sigma} & = & -\sin \beta \cos \psi, \\
Q_1^{\sigma} &=&\frac{1}{2} \sin \theta \left(
- \cos \psi \cos 2 \alpha \sin 2 \beta
- 2 \sin \psi \sin 2 \alpha \sin \beta \right) \nonumber\\
&+& \frac{1}{2} \cos \theta \left[
\sin 2 \psi \cos 2 \alpha (1+\cos^2 \beta)
- 2 \cos 2 \psi \sin 2 \alpha \cos \beta
\right],\\
Q_2^{\sigma} &=&
\frac{\sqrt{3}}{2} \sin \theta \cos \psi \sin 2 \beta
+ \frac{\sqrt{3}}{2} \cos \theta \sin 2 \psi \sin^2 \beta, \\
Q_3^{\sigma} &=&
\sin \theta ( -\cos \psi \sin \alpha \cos 2 \beta
+\sin \psi \cos \alpha \cos \beta ) \nonumber\\
&+& \frac{1}{2} \cos \theta (
-\sin 2 \psi \sin \alpha \sin 2\beta
- 2 \cos 2 \psi \cos \alpha \sin \beta ), \\
Q_4^{\sigma} &=&
\sin \theta ( -\cos \psi \cos \alpha \cos 2 \beta
-\sin \psi \sin \alpha \cos \beta ) \nonumber\\
&+& \frac{1}{2} \cos \theta (
-\sin 2 \psi \cos \alpha \sin 2\beta
+ 2 \cos 2 \psi \sin \alpha \sin \beta ), \\
Q_5^{\sigma} &=&
\frac{1}{2} \sin \theta \left(
- \cos \psi \sin 2 \alpha \sin 2 \beta
+ 2 \sin \psi \cos 2 \alpha \sin \beta \right) \nonumber\\
&+& \frac{1}{2} \cos \theta \left[
\sin 2 \psi \sin 2 \alpha (1+\cos^2 \beta)
+ 2 \cos 2 \psi \cos 2 \alpha \cos \beta
\right]. 
\end{eqnarray}
For $(004\ell+2)$, putting $\alpha=\beta=0$, we have
$P_{x}^{\sigma} = \cos \psi,
 P_{y}^{\sigma} = -\sin \psi, 
 P_{z}^{\sigma} = 0$, and 
$Q_{1}^{\sigma} = \sin 2\psi \cos \theta,
 Q_{2}^{\sigma} = 0,
 Q_{3}^{\sigma} = \sin \psi \sin \theta,
 Q_{4}^{\sigma} =-\cos \psi \sin \theta,
 Q_{5}^{\sigma} = \cos 2\psi \cos \theta $.
For $(\ell\ell\ell)$, putting $\alpha=\pi/4$ and
$\beta=\sin^{-1}\sqrt{2/3}$,
we have
\begin{eqnarray}
 P_{x}^{\sigma} &=& \sqrt{\frac{1}{6}}\left[\cos \psi 
                 + \sqrt{3}\sin \psi\right], \\
 P_{y}^{\sigma} &=& \sqrt{\frac{1}{6}}\left[\cos \psi 
                 - \sqrt{3}\sin \psi\right], \\
 P_{z}^{\sigma} &=& -\sqrt{\frac{2}{3}}\cos \psi ,\\
 Q_{1}^{\sigma} &=& \sqrt{\frac{1}{3}}\left[-\cos 2\psi \cos \theta
                 - \sqrt{2}\sin \psi \sin \theta \right], \\
 Q_{2}^{\sigma} &=& \sqrt{\frac{1}{3}}\left[\sin 2\psi \cos \theta
                 + \sqrt{2}\cos \psi \sin \theta \right], \\
 Q_{3}^{\sigma} &=& \frac{1}{3}\sqrt{\frac{1}{2}}\bigl[
                 - \sqrt{6}\cos 2\psi \cos \theta 
		 - \sqrt{2}\sin 2\psi \cos \theta \nonumber \\
                &+& \cos \psi \sin \theta + \sqrt{3}\sin \psi \sin \theta 
		\bigr], \\
 Q_{4}^{\sigma} &=& \frac{1}{3}\sqrt{\frac{1}{2}}\bigl[
                  \sqrt{6}\cos 2\psi \cos \theta 
		 - \sqrt{2}\sin 2\psi \cos \theta \nonumber \\
                &+& \cos \psi \sin \theta - \sqrt{3}\sin \psi \sin \theta 
		\bigr], \\
 Q_{5}^{\sigma} &=& \frac{\sqrt{2}}{3}\bigl[\sqrt{2}\sin 2\psi \cos \theta 
		 - \cos \psi \sin \theta \bigr].
\end{eqnarray}
The expressions for the scattered photon are obtained by replacing $\theta$
with $-\theta$ in the above expressions.


\begin{thebibliography}{99}
\bibitem{Com3}
   In many literatures, the scattering vector is defined by
   ${\bf G}={\bf k}-{\bf k}'$, which is different from
   our definition ${\bf G}={\bf k}'-{\bf k}$.
   The difference in definition is discernible for systems 
   without centrosymmetry.
\end{thebibliography}
\end{document}